\documentclass[showpacs,amsmath,amssymb,superscriptaddress,showkeys,nofootinbib]{revtex4}

\usepackage{graphicx}
\usepackage{dcolumn}
\usepackage{bm}

\newcommand{\Ref}[1]{(\ref{#1})}
\newcommand{\arxiv}[1]{[arXiv: gr-qc/#1]}
\newcommand{\cqg}{Class. Quantum Grav.\ }

\begin{document}

\title{Non-minimal Wu-Yang wormhole}

\author{A. B. Balakin}
\email{Alexander.Balakin@ksu.ru} \affiliation{Department of
General Relativity and Gravitation, Kazan State University,
Kremlevskaya str. 18, Kazan 420008, Russia}
\author{S. V. Sushkov}
\email{sergey_sushkov@mail.ru; Sergey.Sushkov@ksu.ru} \affiliation{Department of General
Relativity and Gravitation, Kazan State University, Kremlevskaya
str. 18, Kazan 420008, Russia} \affiliation{Department of Mathematics,
Tatar State University of Humanities and Education,\\ Tatarstan str. 2, Kazan 420021, Russia}%
\author{A. E. Zayats}%
\email{Alexei.Zayats@ksu.ru} \affiliation{Department of General
Relativity and Gravitation, Kazan State University, Kremlevskaya
str. 18, Kazan 420008,
Russia}%


\begin{abstract}
We discuss exact solutions of three-parameter non-minimal
Einstein-Yang-Mills model, which describe the wormholes of a new
type. These wormholes are considered to be supported by
SU(2)-symmetric Yang-Mills field, non-minimally coupled to
gravity, the Wu-Yang ansatz for the gauge field being used. We
distinguish between regular solutions, describing traversable
non-minimal Wu-Yang wormholes, and black wormholes possessing one
or two event horizons. The relation between the asymptotic mass of
the regular traversable Wu-Yang wormhole and its throat radius is
analysed.

\end{abstract}

\pacs{04.20.Jb, 14.80.Hv, 04.20.Gz}

\keywords{Einstein-Yang-Mills theory, non-minimal coupling,
traversable wormhole}

\maketitle

\section{Introduction}

Wormholes are topological handles in spacetime linking widely
separated regions of a single universe, or ``bridges'' joining two
different spacetimes \cite{VisserBook}. Recent interest in these
configurations has been initiated by Morris and Thorne
\cite{MorTho}. These authors constructed and investigated a class
of objects they referred to as ``traversable wormholes''.

The central feature of wormhole physics is the fact that
traversable wormholes are accompanied by an unavoidable violation
of the null energy condition, i.e., the matter threading the
wormhole's throat has to be possessed of ``exotic'' properties
\cite{MorTho,HocVis}. The known classical matter does satisfy the
usual energy conditions, hence physical models providing the
existence of wormholes must include hypothetical forms of matter.
Various models of such kind have been considered in the
literature, among them scalar fields \cite{scalarfields}; wormhole
solutions in semi-classical gravity \cite{semiclas}; solutions in
Brans-Dicke theory \cite{Nan-etal}; wormholes on the brane
\cite{wormholeonbrane}; wormholes supported by matter with an
exotic equation of state, namely, phantom energy \cite{phantom},
the generalized Chaplygin gas \cite{chaplygin}, tachyon matter
\cite{tachyon}, etc \cite{footnote,review}.

The electromagnetic field and the non-Abelian gauge field can also
be considered as sources for wormholes when they satisfy the
necessary unusual energy conditions. Such possibility can in
principle appear if one considers nonlinear electrodynamics
\cite{AreLob} or takes into account the non-minimal coupling of
gravity with vector-type fields, i.e., with the non-Abelian
Yang-Mills field, or Maxwell field. The non-minimal
Einstein-Maxwell theory has been elaborated in detail in both
linear (see, e.g., \cite{FaraR,HehlObukhov} for a review) and
non-linear (see, \cite{BL05}) versions. As for the non-minimal
Einstein-Yang-Mills theory, two concepts to derive the master
equations are known. The first one is a dimensional reduction of
the Gauss-Bonnet action \cite{MH}, this model contains one
coupling parameter. The second concept is a non-Abelian
generalization of the non-minimal non-linear Einstein-Maxwell
theory \cite{BL05}. We will follow the latter approach and
construct a three-parameter non-minimal model being linear in the
curvature by analogy with the well-known model proposed by
Drummond and Hathrell for linear electrodynamics \cite{Drum}.
Three coupling constants $q_1$, $q_2$, and $q_3$ of the model are
shown to introduce a new specific radius associated with the
radius $a$ of the wormhole throat.

In this work we focus on the example of exact solution of the
non-minimal three-parameter EYM model describing the wormhole of a
new type, namely, {\em non-minimal wormhole}. It can also be
indicated as {\em non-minimal Wu-Yang wormhole}, since the
solution of the non-minimally extended Yang-Mills subsystem of the
total self-consistent EYM system of equations is the direct analog
of the Wu-Yang monopole \cite{WuYang}.

The paper is organized as follows. In Sec. \ref{formalism} we
briefly describe the formalism of three-parameter non-minimal
Einstein-Yang-Mills model. In Sec. \ref{exact}, Subsect. A we
adapt this model for the case of static spherically symmetric
field configuration, present the exact solution of the Wu-Yang
type to the gauge field equations and formulate two key equations
for two metric functions $\sigma(r)$ and $N(r)$. In Subsect. B, C
and D we discuss the details of the three-parameter family of
exact solutions for the function $\sigma(r)$. Sec. \ref{travWH} is
devoted to the analysis of the solution describing the non-minimal
Wu-Yang wormhole. Conclusions are formulated in the last section.

\section{Non-minimal Einstein-Yang-Mills model}\label{formalism}

The action of the three-parameter non-minimal Einstein-Yang-Mills
model has the form\footnote{Hereafter we use the units
$c=G=\hbar=1$.}
\begin{equation}\label{action}
S_{{\rm NMEYM}} = \int d^4 x \sqrt{-g}\
\left(\frac{1}{8\pi}R+\frac{1}{2}F^{(a)}_{ik}
F^{ik(a)}+\frac{1}{2} \chi^{ikmn} F^{(a)}_{ik}
F^{(a)}_{mn}\right)\,,
\end{equation}
where $g = {\rm det}(g_{ik})$ is the determinant of a metric
tensor $g_{ik}$, and $R$ is the Ricci scalar. The Latin indices
without parentheses run from 0 to 3, the summation with respect to
the repeated group indices $(a)$ is implied. The tensor
$\chi^{ikmn}$, indicated in \cite{BL05} as non-minimal
susceptibility tensor, is defined as follows:
\begin{equation}
\chi^{ikmn} \equiv \frac{q_1}{2}R\,(g^{im}g^{kn}-g^{in}g^{km}) +
\frac{q_2}{2}(R^{im}g^{kn} - R^{in}g^{km} + R^{kn}g^{im}
-R^{km}g^{in}) + q_3 R^{ikmn} \,.
\end{equation}
Here $R^{ik}$ and $R^{ikmn}$ are the Ricci and Riemann tensors,
respectively, and $q_1$, $q_2$, $q_3$ are the phenomeno\-logi\-cal
parameters describing the non-minimal coupling of the Yang-Mills
and gravitational fields. Following \cite{Rubakov}, we consider
the Yang-Mills field, ${\bf F}_{mn}$, to take the values in the
Lie algebra of the gauge group $SU(2)$:
\begin{equation}
{\bf F}_{mn} = - i {\cal G}  F^{(a)}_{mn} {\bf t}_{(a)} \,, \quad
{\bf A}_m = - i {\cal G}  A^{(a)}_m {\bf t}_{(a)} \,.
\label{represent}
\end{equation}
Here ${\bf t}_{(a)}$ are Hermitian traceless generators of $SU(2)$
group, ${\cal G}$ is a constant of gauge interaction, and the group
index $(a)$ runs from 1 to 3. The generators ${\bf t}_{(a)}$ satisfy
the commutation relations:
\begin{equation}
    [{\bf t}_{(a)},{\bf t}_{(b)}]=i\,\varepsilon_{(a)(b)(c)}{\bf
    t}_{(c)}\,,
\end{equation}
where $\varepsilon_{(a)(b)(c)}$ is the completely antisymmetric
symbol with $\varepsilon_{(1)(2)(3)}=1$. The Yang-Mills field
potential, $\mathbf{A}_i$, and strength field, $\mathbf{F}_{ik}$,
are coupled by the relation
\begin{equation}\label{strength}
\mathbf{F}_{ik}=\partial_i\mathbf{A}_k-\partial_k\mathbf{A}_i+
\left[\mathbf{A}_i\,,\mathbf{A}_k\right] \,, 
\end{equation}
which guarantees that the equation
\begin{equation}
\hat{D}_l \mathbf{F}_{ik} + \hat{D}_k \mathbf{F}_{li} + \hat{D}_i
\mathbf{F}_{kl} = 0 \label{1Fik}
\end{equation}
turns into identity. Here the symbol $\hat{D}_k$ denotes the gauge
invariant derivative
\begin{equation}
\hat{D}_i \equiv \nabla_i + \left[{\bf A}_i \,,  \ \ \  \right]
\,, \label{11Fik}
\end{equation}
and $\nabla _m$ is a covariant spacetime derivative.

The variation of the action \Ref{action} with respect to
Yang-Mills potential $A^{(a)}_i$ yields
\begin{equation}
\hat{D}_k {\bf H}^{ik} \equiv \nabla_k {\bf H}^{ik}+\left[{\bf
A}_k,{\bf H}^{ik}\right] = 0 \,, \quad {\bf H}^{ik} = {\bf F}^{ik}
+ \chi^{ikmn} {\bf F}_{mn} \,. \label{YM}
\end{equation}
The tensor ${\bf H}^{ik}$ is a non-Abelian analog of the induction
tensor known in the electrodynamics \cite{Maugin}, and thus
$\chi^{ikmn}$ can be considered as a non-minimal susceptibility
tensor \cite{BL05}. The variation of the action with respect to
the metric $g_{ik}$ yields
\begin{equation}
R_{ik} - \frac{1}{2} R \ g_{ik} = 8\pi\,T^{({\rm eff})}_{ik} \,.
\label{Ein}
\end{equation}
The effective stress-energy tensor $T^{({\rm eff})}_{ik}$ can be
divided into four parts:
\begin{equation}
T^{({\rm eff})}_{ik} =  T^{(YM)}_{ik} + q_1 T^{(I)}_{ik} + q_2
T^{(II)}_{ik} + q_3 T^{(III)}_{ik} \,. \label{Tdecomp}
\end{equation}
The first term $T^{(YM)}_{ik}$:
\begin{equation}
T^{(YM)}_{ik} \equiv \frac{1}{4} g_{ik} F^{(a)}_{mn}F^{mn(a)} -
F^{(a)}_{in}F_{k}^{\ n(a)} \,, \label{TYM}
\end{equation}
is a stress-energy tensor of the pure Yang-Mills field. The
definitions of other three tensors relate to the corresponding
coupling constants $q_1$, $q_2$, $q_3$:
\begin{equation}%
T^{(I)}_{ik} = R\,T^{(YM)}_{ik} -  \frac{1}{2} R_{ik}
F^{(a)}_{mn}F^{mn(a)} + \frac{1}{2} \left[ \hat{D}_{i} \hat{D}_{k}
- g_{ik} \hat{D}^l \hat{D}_l \right] \left[F^{(a)}_{mn}F^{mn(a)}
\right] \,, \label{TI}
\end{equation}%

\[%
T^{(II)}_{ik} = -\frac{1}{2}g_{ik}\biggl[\hat{D}_{m}
\hat{D}_{l}\left(F^{mn(a)}F^{l\ (a)}_{\ n}\right)-R_{lm}F^{mn (a)}
F^{l\ (a)}_{\ n} \biggr] \]%
\[{}- F^{ln(a)}
\left(R_{il}F^{(a)}_{kn} +
R_{kl}F^{(a)}_{in}\right)-R^{mn}F^{(a)}_{im} F_{kn}^{(a)} -
\frac{1}{2} \hat{D}^m \hat{D}_m \left(F^{(a)}_{in} F_{k}^{ \
n(a)}\right)\]%
\begin{equation}%
\quad{}+\frac{1}{2}\hat{D}_l \left[ \hat{D}_i \left(
F^{(a)}_{kn}F^{ln(a)} \right) + \hat{D}_k
\left(F^{(a)}_{in}F^{ln(a)} \right) \right] \,, \label{TII}
\end{equation}%

\[
T^{(III)}_{ik} = \frac{1}{4}g_{ik}
R^{mnls}F^{(a)}_{mn}F_{ls}^{(a)}- \frac{3}{4} F^{ls(a)}
\left(F_{i}^{\ n(a)} R_{knls} +
F_{k}^{\ n(a)}R_{inls}\right) \]%
\begin{equation}%
\quad {}-\frac{1}{2}\hat{D}_{m} \hat{D}_{n} \left[ F_{i}^{ \ n
(a)}F_{k}^{ \ m(a)} + F_{k}^{ \ n(a)} F_{i}^{ \ m(a)} \right] \,.
\label{TIII}
\end{equation}%
The tensor $T^{({\rm eff})}_{ik}$ satisfies the conservation law
$\nabla^k T^{({\rm eff})}_{ik} =0$. The self-consistent system of
equations (\ref{YM}) and (\ref{Ein}) with
(\ref{Tdecomp})-(\ref{TIII}) is a direct non-Abelian
generalization of the three-parameter non-minimal Einstein-Maxwell
model discussed in \cite{BL05}. This system can also be considered
as one of the variants of a non-minimal generalization of the
Einstein-Yang-Mills model.

\section{Exact solutions of the static model with spherical symmetry}
\label{exact}

\subsection{Master equations}

Let us take the metric of a static spherically symmetric spacetime
in the form to be especially convenient for studying a wormhole
geometry:
\begin{equation}\label{metrica}
ds^2=\sigma^2Ndt^2-\frac{dr^2}{N}-\left(r^2+a^2\right)\left(d\theta^2+\sin^2\theta
d\varphi^2\right),
\end{equation}
where the metric functions $\sigma$ and $N$ depend only on $r$.
The properties of {\em traversable} wormholes dictate some
additional requirements for the metric \Ref{metrica}, which were
in great detail discussed in \cite{MorTho,VisserBook}. In
particular, we note that

\begin{enumerate}

\item[(i)] the radial coordinate $r$ runs from $-\infty$ to
$+\infty$. Two asymptotical regions $r=-\infty$ and $r=+\infty$
are connected by the wormhole's throat which has the radius $a$
and is located at $r=0$.

\item[(ii)] Since the spacetime of a traversable wormhole has
neither singularities nor event horizons, the metric components
$g_{tt}=\sigma^2 N$ and $-g_{rr}=1/N$ should be  regular and
positive everywhere. Note that, in particular, this means that
$N(r)$ is positive defined, both $\sigma(r)$ and $N(r)$ are
finite, and neither $\sigma(r)$ nor $N(r)$ can take zero values.

\item[(iii)] In addition, one may demand the asymptotical flatness of the
wormhole spacetime at $r=\pm\infty$. This is guaranteed provided
the following boundary conditions for the functions $\sigma$ and
$N$ are satisfied:
\begin{equation}\label{asymp}
\sigma^2\left(\pm\infty\right)=1 \,, \quad
N\left(\pm\infty\right)=1 .
\end{equation}
\end{enumerate}
Below we will search for solutions of the non-minimal
Einstein-Yang-Mills model, which satisfy the listed requirements.

The non-minimal Yang-Mills equations (\ref{YM}) are satisfied
identically, when the gauge field is parameterized as
\cite{RebbiRossi,BaZa07}
\begin{equation}\label{1}
\mathbf{A}_{0}=\mathbf{A}_{r}=0\,,\quad\mathbf{A}_{\theta}=i\mathbf{t}_{\varphi}\,,\quad
\mathbf{A}_{\varphi}=-{i\nu}\sin{\theta}\;\mathbf{t}_{\theta} \,,
\end{equation}
which is known to be the so-called Wu-Yang monopole solution
\cite{WuYang}. The parameter $\nu$ is a non-vanishing integer,
${\bf t}_r$, ${\bf t}_{\theta}$ and ${\bf t}_{\varphi}$ are the
position-dependent generators of the SU(2) group:
\[%
{\bf t}_r=\cos{\nu\varphi}\sin{\theta}\;{\bf
t}_{(1)}+\sin{\nu\varphi}\sin{\theta}\;{\bf
t}_{(2)}+\cos{\theta}\;{\bf t}_{(3)},
\]%
\begin{equation}%
{\bf t}_{\theta}=\partial_{\theta}{\bf t}_r,\qquad {\bf
t}_{\varphi}=\frac {1}{\nu\sin{\theta}}\ \partial_{\varphi}{\bf
t}_r, \label{deS5}
\end{equation}%
which satisfy the following commutation rules
\begin{equation}%
\left[{\bf t}_{r},{\bf t}_{\theta}\right]=i\,{\bf
t}_{\varphi},\quad \left[{\bf t}_{\theta},{\bf
t}_{\varphi}\right]=i\,{\bf t}_{r}, \quad \left[{\bf
t}_{\varphi},{\bf t}_{r}\right]=i\,{\bf t}_{\theta}.\label{deS6}
\end{equation}%
The field strength tensor $\mathbf{F}_{ik}$ has only one
non-vanishing component
\begin{equation}\label{2}
{\bf F}_{\theta\varphi}={i\nu}\sin\theta\,{\bf
t}_{r}\,. 
\end{equation}
Since the effective energy-momentum tensor $T^{({\rm eff})}_{ik}$
is divergence-free, the Einstein equations for the spherical
symmetric metric (\ref{metrica}) are known to be effectively
reduced to the two key equations, say, for equations with $i=k=0$
and $i=k=r$. The components $G_0^{\,0}$ and $G_r^{\,r}$ of the
Einstein tensor $G_i^{\,k}=R_i^k-\frac{1}{2}\delta_i^kR$ are
\begin{equation}
G_0^{\,0}=\frac{1-rN'-N}{r^2+a^2}-\frac{N\, a^2}{(r^2+a^2)^2} \,,
\end{equation}
\begin{equation}
G_r^{\,r}=\frac{1-rN'-N}{r^2+a^2}-\frac{2rN\sigma'}{\sigma(r^2+a^2)}+\frac{a^2N}{(r^2+a^2)^2}
\,.
\end{equation}
The corresponding components of the effective energy-momentum
tensor (see (\ref{TYM})-(\ref{TIII})) take the form
\begin{equation}
T_0^{0}=\frac{\nu^2}{{\cal
G}^2}\left(\frac{1}{2(r^2+a^2)^2}-\frac{2q_1Na^2}{(r^2+a^2)^4}-\frac{q_1N'r+q_1+q_2+q_3}{(r^2+a^2)^3}+
\frac{(13q_1+4q_2+q_3)Nr^2}{(r^2+a^2)^4}\right) \,,
\end{equation}
\begin{equation}
T_r^{r}=\frac{\nu^2}{{\cal
G}^2}\left[\frac{1}{2(r^2+a^2)^2}-\frac{1}{(r^2+a^2)^3}\left(q_1N'r+q_1+q_2+q_3+
\frac{2q_1Nr\sigma'}{\sigma}\right)
-\frac{(7q_1+4q_2+q_3)Nr^2}{(r^2+a^2)^4}\right] \,.
\end{equation}
The difference
$G_0^{\,0}{-}G_r^{\,r}{=}8\pi\left(T_0^0{-}T_r^r\right)$ of these
equations can be transformed into the decoupled equation for
$\sigma(r)$ only
\begin{equation}
r  \left[1- \frac{\kappa q_1}{(r^2+a^2)^2} \right] \
\frac{\sigma^{\prime}}{\sigma} =
\frac{a^2}{(r^2+a^2)}+\frac{\kappa}{(r^2+a^2)^3}\left[(10q_1+4q_2+q_3)r^2-q_1a^2\right]
\,, \label{sisi}
\end{equation}
where $\kappa$ is a new charge parameter with the dimensionality
of area, $\kappa=8\pi\nu^2/{{\cal G}^2}$. Solving this equation
one can find the function $\sigma(r)$ in the explicit form for
arbitrary values of parameters $q_1$, $q_2$ and $q_3$:
\begin{equation}\label{sigma}
    \sigma(r)=\frac{r}{\sqrt{r^2+a^2}}\left[1-\frac{\kappa
q_1}{(r^2+a^2)^2}\right]^{\frac{10q_1+4q_2+q_3}{4q_1}} \,.
\end{equation}
Excluding the function $\sigma$ from the Einstein equation with
$i=k=0$, we obtain the equation for $N(r)$ only:
$$
r \left[1 - \frac{\kappa  q_1}{(r^2+a^2)^2} \right] N^{\prime} + N
\left[1 + \frac{a^2}{(r^2+a^2)} + \frac{\kappa
(13q_1+4q_2+q_3)}{(r^2+a^2)^2} -  \frac{\kappa a^2
(15q_1+4q_2+q_3)}{(r^2+a^2)^3} \right]=
$$
\begin{equation}
{}= 1 - \frac{\kappa}{2 (r^2+a^2)}+
\frac{\kappa(q_1+q_2+q_3)}{(r^2+a^2)^2} \,. \label{nana}
\end{equation}
The solution of Eq. (\ref{nana}) can be clearly represented in
quadratures. Note that the coefficient $\Theta(r) \equiv r
\left[1-\kappa q_1(r^2+a^2)^{-2}\right]$ in front of the first
(highest) derivative in both differential equations (\ref{sisi})
and (\ref{nana}) can take, in principle, zero values depending on
the sign of the guiding parameter $q_1$. Thus, searching for the
solutions of these equations, we have to distinguish three
qualitatively different cases $q_1<0$, $q_1=0$ and $q_1>0$.

\subsection{The case $q_1<0$}

For negative $q_1$ the coefficient $\Theta(r)$ has only one root,
$r=0$. In this case, the function $\sigma(r)$ takes the form
\begin{equation}
\sigma(r)=\frac{r}{\sqrt{r^2+a^2}}\left[1+\frac{\kappa
|q_1|}{(r^2+a^2)^2}\right]^{\frac{10|q_1|-4q_2-q_3}{4|q_1|}} \,.
\label{sisol}
\end{equation}
Notice that $\sigma(r)$ given by Eq. \Ref{sisol} turns into zero
at $r=0$, i.e., $\sigma(0)=0$. This violates the condition (ii),
and so this solution cannot describe a traversable wormhole. Note
also that $\sqrt{-g(0)}=0$, where $g=-\sigma^2
(r^2+a^2)^2\sin^2\theta$ is the determinant of $g_{ik}$. This
means that the chosen coordinate system is ill-defined at $r=0$,
and well-defined only in the range $(0,+\infty)$ (or,
equivalently, in $(-\infty,0)$).

\subsection{The case $q_1=0$}

In this case $\Theta(r)=r$, and so $r=0$ is the only root of
$\Theta(r)$ as in the case $q_1<0$. The solution for $\sigma(r)$
transforms now into
\begin{equation}
\sigma(r) = \frac{r}{\sqrt{r^2+a^2}}\exp \left\{ - \frac{\kappa
(4q_2+q_3)}{(r^2+a^2)^2}\right\}\,. \label{1sisol}
\end{equation}
The function $\sigma(r)$ given by Eq. \Ref{1sisol} turns into zero
at $r=0$, i.e., $\sigma(0)=0$. This means that the case $q_1=0$ does
not admit the existence of traversable wormholes.

\subsection{The case $q_1>0$}

For positive $q_1$ the number of real roots of $\Theta(r)$ depends
on the value $\beta \equiv (\kappa q_1)^{1/4}$. In case $\beta
> a$, $\Theta(r)$ has three real roots, namely, $r=0$, $r=\pm
r^*$, where
\begin{equation}\label{singa}
r^* \equiv \sqrt{\beta^2 - a^2} \,.
\end{equation}
For $\beta = a$ the roots $r= \pm r^*$ coincide with $r=0$, and for
$\beta < a$ one has only one real root $r=0$. Below we consider each
case separately.

\vskip6pt I. {$\beta < a$.} Rewrite the solution \Ref{sigma} as
follows
\begin{equation}\label{sig1}
    \sigma(r)=\frac{r}{\sqrt{r^2+a^2}}\left[\frac{(r^2+a^2-\beta^2)
    (r^2+a^2+\beta^2)}{(r^2+a^2)^2}\right]^{\frac{10q_1+4q_2+q_3}{4q_1}} \,.
\end{equation}
It is clear that due to the condition $\beta < a$ the expression
in square brackets in Eq. \Ref{sig1} is positive for all $r$.
Thus, for all values of the power parameter
$(10q_1+4q_2+q_3)/4q_1$ the sign of the function $\sigma(r)$
inherits the sign of $r$, and turns into zero at $r=0$, i.e.,
$\sigma(0)=0$. As in previous cases, this means that traversable
wormholes do not exist.

\vskip6pt II. $\beta > a$. In this case the expression in square
brackets in Eq. \Ref{sig1} vanishes, when
$r=r^*\equiv(\beta^2-a^2)^{1/2}$. Now, depending on the sign of
the power parameter $(10q_1+4q_2+q_3)/4q_1$, the solution
$\sigma(r)$ turns into zero or tends to infinity at $r^*$. When
$10q_1+4q_2+q_3=0$, one has again $\sigma(0)=0$. Thus, the case
$\beta > a$ also does not admit traversable wormholes.

\vskip6pt III. $\beta = a$  (or, equivalently, $q_1=a^4/\kappa$).
 It will be convenient to rewrite the solution \Ref{sig1} for
$\sigma(r)$ in the following form:
\begin{equation}
\sigma(r) = \sqrt{\frac{r^2+a^2}{r^2+2a^2}} \ \ \left[ \frac{r^2
(r^2+2a^2)}{(r^2+a^2)^2} \right]^{\frac{12 q_1+4q_2+q_3}{4q_1}} \,.
\label{reg1}
\end{equation}
Now the critical point of interest is $r=0$. The behavior of
$\sigma(r)$ near $r=0$ essentially depends on the sign of the new
power parameter, namely, $12 q_1+4q_2+q_3$. In particular, for $12
q_1+4q_2+q_3>0$ one has $\sigma(0)=0$, and for $12 q_1+4q_2+q_3<0$
one has $\sigma(0)=\infty$. Such behavior of $\sigma(r)$ excludes
traversable wormholes. Consider the last particular case, when
this parameter vanishes, $12 q_1+4q_2+q_3=0$. Now we obtain
\begin{equation}\label{sigreg}
\sigma(r) = \sqrt{\frac{r^2+a^2}{r^2+2a^2}} \,.
\end{equation}
The function $\sigma(r)$ given by Eq. \Ref{sigreg} is regular and
positive in the whole interval $(-\infty,+\infty)$, moreover,
$\sigma(\pm\infty)=1$. Thus, $\sigma(r)$ given by Eq. \Ref{sigreg}
satisfies the necessary conditions (i-iii) and the corresponding
field configuration can be considered as a candidate in searching
for traversable wormholes. In the next section we will complete
the solution for $\sigma(r)$ by the solution for $N(r)$ and
discuss the properties of the non-minimal Wu-Yang wormhole
solution.

\section{Non-minimal Wu-Yang wormhole}\label{travWH}

In this section we consider in more details the special case
corresponding to the following choice of the non-minimal coupling
parameters $q_1$, $q_2$, $q_3$:
\begin{equation}\label{qqq1}
q_1=a^4/\kappa,\quad 12q_1+4q_2+q_3=0.
\end{equation}
Then, the equation (\ref{nana}) can be easily integrated in the
quadratures to give
\begin{equation} \label{N}
N(r)=\frac{(r^2+a^2)^{3/2}}{r^3\,\sqrt{r^2+2a^2}} \left\{ C +
\int\limits_0^r \frac{dx}{(x^2+a^2)^{3/2}\sqrt{x^2+2a^2}}
\left[x^4 + 2x^2 \left(a^2-\frac{\kappa}{4}\right) -
\left(10a^4+\frac{\kappa a^2}{2} + 3\kappa q_2 \right)\right]
\right\},
\end{equation}
where $C$ is a constant of integration. Note that for arbitrary
values of $a$, $q_2$, and $C$ the function $N(r)$ given by Eq.
\Ref{N} satisfies the boundary condition $N(\pm\infty)=1$. Near
$r=0$ the solution $N(r)$ is, generally speaking, divergent. Such
behavior of $N(r)$ is unsuitable for description of traversable
wormholes. However, there are special values of parameters $q_2$
and $C$, namely:
\begin{equation} \label{Nu}
C=0 \,, \quad q_2=-\frac{10a^4}{3\kappa}-\frac{a^2}{6} \,,
\end{equation}
for which the solution \Ref{N} transforms into
\begin{equation}\label{Nreg}
N(r)=\frac{(r^2+a^2)^{3/2}}{r^3\sqrt{r^2+2a^2}} \ J(r) \,,
\end{equation}
where
\begin{equation}
J(r)=\int\limits_0^r\frac{x^2dx}{(x^2+a^2)^{3/2}\sqrt{x^2+2a^2}}\left(x^2
+ 2a^2 -\frac{\kappa}{2} \right) \label{g}
\end{equation}
is a function of $r$ and two guiding parameters, $a$ and $\kappa$.
Note that near $r=0$ the function $N(r)$, given by Eq. \Ref{Nreg},
behaves as
\begin{equation}
N(r) \simeq (3a^2)^{-1}(a^2-\kappa/4)+O(r^2) \,. \label{gd}
\end{equation}
It is seen that $N(r)$ can be positive, negative, or zero at $r=0$
depending on the relation between two parameters: $a$ (the
wormhole throat radius) and $\kappa$ (the charge parameter). It
will be convenient further to use a dimensionless parameter
$\alpha=a\kappa^{-1/2}$. The behavior of $N(r)$ depending on
$\alpha$ is illustrated in the Fig.1.

\begin{figure}[h]
\centerline{\includegraphics{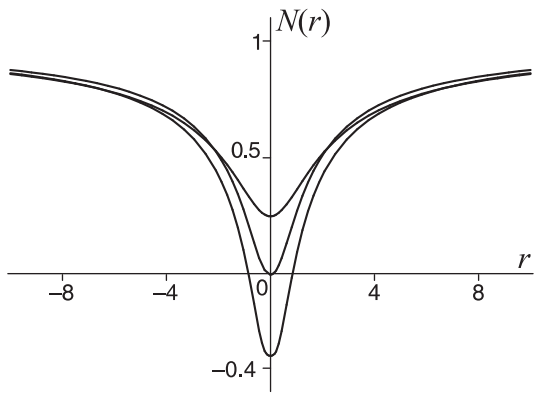}} \caption{Graphs of the
function $N(r)$ given for $\alpha\equiv a\kappa^{-1/2}>1/2$,
$\alpha=1/2$, and $\alpha<1/2$ from up to down,
respectively.}\label{FigN}
\end{figure}

Taking into account the relations $g_{tt}=\sigma^2N$ and
$-g_{rr}=1/N$ and using the solutions \Ref{sigreg} and \Ref{Nreg}
for $\sigma(r)$ and $N(r)$ we finally obtain the following metric,
which presents the new exact solution of the non-minimally
extended Einstein-Yang-Mills equations:
\begin{equation}\label{finalmetric}
ds^2=\frac{(r^2+a^2)^{5/2}}{r^3(r^2+2a^2)^{3/2}}\,J(r)dt^2-
\frac{r^3(r^2+2a^2)^{1/2}}{(r^2+a^2)^{3/2}}\,\frac{dr^2}{J(r)}-
(r^2+a^2)\,(d\theta^2+\sin^2\theta d\varphi^2) \,.
\end{equation}
This metric describes a regular (i.e., without singularities)
spacetime containing two asymptotically flat regions $r=\pm\infty$
connected by a throat located at $r=0$. Thus, the metric
\Ref{finalmetric} describes a wormhole, which we will hereafter
call as a {\em non-minimal Wu-Yang wormhole}.

The spacetime structure of the Wu-Yang wormhole essentially
depends on the value of the dimensionless parameter
$\alpha=a\kappa^{-1/2}$. We note that for $\alpha>1/2$ the
function $N(r)$ is positive defined (see Fig. 1), and so the
metric components $g_{tt}=\sigma^2N$ and $-g_{rr}=1/N$ are finite
and positive in the whole region $(-\infty,+\infty)$. This means
that the spacetime has no event horizons, thus in this case the
Wu-Yang wormhole is {\em traversable}.

In case $\alpha<1/2$ the function $N(r)$ changes the sign. It is
positive for $|r|>r_h$, negative for $|r|<r_h$, and zero at
$|r|=r_h$, i.e., $N(\pm r_h)=0$ ($r_h$ is some parameter, which
can be easily found numerically for every $\alpha<1/2$). In the
vicinity of $|r|=r_h$ one has $g_{tt}\sim(r-r_h)$ and
$g_{rr}\sim(r-r_h)^{-1}$. This means that the points $|r|=r_h$ are
nothing but two event horizons of Schwarzschild-like type in the
wormhole spacetime, and $r_h$ is the radius of horizons. In the
accepted nomenclature, the regions $|r|>r_h$ with $N(r)>0$ and
$|r|<r_h$ with $N(r)<0$ are R- and T-regions, respectively. Thus,
in the case $\alpha<1/2$ the throat of Wu-Yang wormhole turns out
to be hidden in the T-region behind the horizons. Such a wormhole
is {\rm non-traversable} from the point of view of a distant
observer. By analogy with black holes one may call such objects as
{\em black wormholes}.

Note that for $\alpha=1/2$ two event horizons $|r|=r_h$ merge with
each other and form an event horizon located at the wormhole's
throat $r=0$. Now, in the vicinity of $r=0$ one has $g_{tt}\sim
r^2$ and $g_{rr}\sim r^{-2}$, and this means that $r=0$ is an
extremal horizon. In this case the T-region is absent, and the
event horizon divides two R-regions.

\begin{figure}[h]
\centerline{\includegraphics{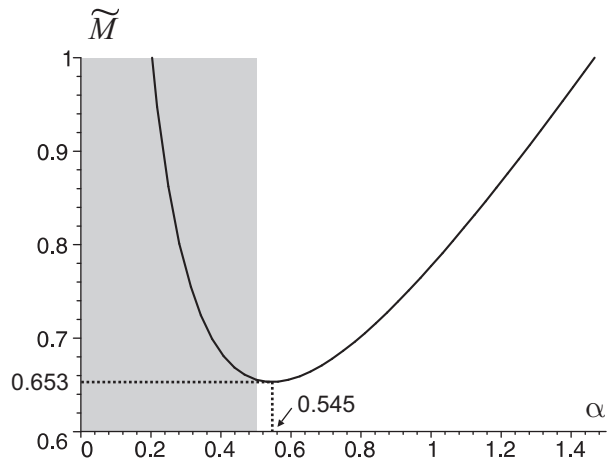}} \caption{Wormhole mass
$\widetilde M(\alpha)$. The shaded region corresponds to
$\alpha<1/2$.}\label{FigM}
\end{figure}

Now let us discuss a formula for an asymptotic mass of the Wu-Yang
wormhole measured by a distant observer. A mass of a static
spherically symmetric configuration is defined as
$M=\frac12\lim_{r\to\pm\infty}\big\{|r|\,(1-g_{tt}(r))\big\}$.
Using the metric \Ref{finalmetric} we can obtain after some
algebra the following expression for the mass of the non-minimal
Wu-Yang wormhole:
\begin{equation} \label{mass}
\frac{M}{\kappa^{1/2}}\equiv\widetilde M(\alpha) = \frac{\pi
\sqrt{2\pi}}{\Gamma^2\!\left(\frac14\right)}
\left(\alpha-\frac{1}{4\alpha}\right)+\frac{\Gamma^2\!\left(\frac14\right)}{16\sqrt{2\pi}}
\,\frac1{\alpha} \,,
\end{equation}
where $\Gamma(z)$ is gamma function, and $\alpha=a\kappa^{-1/2}$
and $\widetilde M=M\kappa^{-1/2}$ are dimensionless quantities.
The graph of $\widetilde M(\alpha)$ is given in Fig. \ref{FigM}.
It is worth to note that $\widetilde M(\alpha)$ is positive
defined, $\widetilde M(\alpha)>0$. Moreover, the function
$\widetilde M(\alpha)$ has a minimum $\widetilde M_{\rm
min}\approx 0.653$ at $\alpha=\alpha_{\rm min}\approx 0.545$.

\section{Conclusions}

In this paper we have considered the non-minimally extended
Einstein-Yang-Mills model given by the action \Ref{action}. The
model contains three phenomenological parameters $q_1$, $q_2$ and
$q_3$, which determine the non-minimal coupling of the Yang-Mills
and gravitational fields. In the framework of this model we have
studied static spherically symmetric configurations with the
Yang-Mills field possessing the SU(2) symmetry. Basing on the
Wu-Yang ansatz for the gauge field we have obtained a
three-parameter family of the explicit exact solutions to the
non-linear Einstein-Yang-Mills equations. Only one solution from
this family is regular and belongs to the class of wormhole
spacetimes. We have denoted this solution as a {\em non-minimal
Wu-Yang wormhole} (see Eq. \Ref{finalmetric}). Let us emphasize
some of its properties.

\begin{enumerate}

\item

The non-minimal Wu-Yang wormhole corresponds to the specific
choice of coupling parameters $q_1$, $q_2$, $q_3$, namely,
\begin{equation}
q_1=\frac{a^4}{\kappa}\,,\quad q_2=-\frac{10 a^4}{3\kappa}-
\frac{a^2}{6}\,,\quad q_3=\frac{4a^4}{3\kappa}+\frac{2a^2}{3}\,.
\label{qqq}
\end{equation}
Thus, the Wu-Yang wormhole geometry turns out to be completely
determined by two model parameters: the wormhole throat radius
$a$, and the charge parameter $\kappa=8\pi\nu^2/{\cal G}^2$, or,
equivalently, by $a$ and the dimensionless parameter $\alpha
\equiv a \kappa^{-1/2}$. Note that in the minimal limit, when
$q_1=q_2=q_3=0$, the relations \Ref{qqq} yield $a=0$, i.e., this
wormhole does not exist. In other words, the obtained exact
solution is {\em essentially} non-minimal.

\item

The parameter $\alpha$ can be treated as {\em guiding} one.
Indeed, in case $\alpha>1/2$ the spacetime of Wu-Yang wormhole has
no event horizons, and so it is traversable in principle. The
condition $\alpha>1/2$ equivalently reads $a>\frac12\kappa^{1/2}$,
that is the throat's radius $a$ of traversable Wu-Yang wormholes
is necessary greater than $\frac12\kappa^{1/2}$. In case
$\alpha<1/2$ ($a<\frac12\kappa^{1/2}$) the wormhole spacetime
\Ref{finalmetric} possesses two Schwarzschild-type event horizons
at $|r|=r_h$, where $r_h$ is an event horizon radius given by the
equation $\sigma^2 N(r_h)=0$. The presence of event horizons means
the Wu-Yang wormhole is non-traversable from the point of view of
a distant observer. It is worth to note that in this case the
wormhole throat located at $r=0$ turns out to be hidden behind the
horizons. For this reason one can call such objects as {\em black
wormholes}. For the particular value $\alpha=1/2$
($a=\frac12\kappa^{1/2}$) two event horizons merge with each other
and form a single event horizon at the throat $r=0$. Now in the
vicinity of $r=0$ the metric functions behave as $g_{tt}\sim r^2$
and $g_{rr}\sim r^{-2}$, and so the metric \Ref{finalmetric}
behaves near the horizon as the extreme Reissner-Nordstr\"{o}m
metric.

\item

For a distant observer the Wu-Yang wormhole manifests itself
through its asymptotical mass $M$. It is determined by the charge
parameter $\kappa$ and expressed through the wormhole throat
radius $a$ (see Eq. \Ref{mass} and Fig. \ref{FigM}). Is it
possible for the observer to reconstruct the invisible throat
radius using the estimated mass? In principle, yes, but the
procedure is ambiguous, since two values of $a$ correspond to one
appropriate value of the mass.

\item

The important feature is that there exists the {\em lower limit}
for the mass of the non-minimal Wu-Yang wormhole. In other words,
the wormhole mass cannot be less than some minimal value $M_{\rm
min}\approx 0.653\,\kappa^{1/2}$, i.e., $M\ge M_{\rm min}$. To
make estimations we assume that the monopole magnetic charge $\nu$
is equal to one, $\nu=1$, and the square of the constant of gauge
interaction is given by ${\cal G}^2=4\pi\alpha_{\rm em}$, where
$\alpha_{\rm em}=e^2/\hbar c\approx 1/137$ is the fine structure
constant. Then, in the dimensional units we have $M_{\rm
min}\approx 10.8\, M_{\rm pl}$, $a_{\rm min}\approx 9\, L_{\rm
pl}$, where $M_{\rm pl}$ and $L_{\rm pl}$ are the Planck mass and
the Planck length, respectively.

\end{enumerate}

Recently Kirill Bronnikov attracted our attention to the papers
\cite{BroMelDeh,BroDehMel}, where the authors discuss solutions
they refer to as regular black holes. He also emphasized that
black wormholes obtained in our paper represent the kind of
regular black holes.

\begin{acknowledgments}
This work was partially supported by the Deutsche
Forschungsgemeinschaft through the project No. 436RUS113/487/0-5
and the Russian Foundation for Basic Research grant No.
05-02-17344.
\end{acknowledgments}

\end{document}